\documentstyle[prl,aps,draft,twocolumn,epsf]{revtex}
\begin{document}
\draft
\wideabs{
\title{Test of CPT and Lorentz Invariance from Muonium Spectroscopy}
\author{V.W. Hughes, M. Grosse Perdekamp\cite{matthias}, D. Kawall and W. Liu}
\address{Yale University, Department of Physics, New Haven, CT 06520-8121, USA}
\author{K. Jungmann and G. zu Putlitz}
\address{Universit\"at Heidelberg, Physikalisches Institut,
D-69120 Heidelberg, Germany}
\maketitle
\begin{abstract}
Following a suggestion of Kosteleck\a'{y} {\it{et al.}}
we have evaluated a test of CPT and Lorentz invariance from the microwave
spectroscopy of muonium. Hamiltonian terms
beyond the standard model violating CPT and Lorentz invariance would
contribute frequency shifts $\delta\nu_{12}$ and $\delta\nu_{34}$ to 
$\nu_{12}$ and $\nu_{34}$, 
the two transitions involving muon spin flip, which were 
precisely measured in ground state muonium in a strong magnetic field 
of 1.7 T. The shifts would be indicated by  
anti-correlated oscillations in $\nu_{12}$ and $\nu_{34}$ at
the earth's sidereal frequency. No time dependence was found in 
$\nu_{12}$ or $\nu_{34}$ at the level of 20 Hz, limiting 
the size of some CPT and Lorentz violating parameters 
at the level of $2\times10^{-23}$ GeV, representing Planck scale
sensitivity and an order of magnitude improvement in
sensitivity over previous limits for the muon.
\end{abstract}
%
}
%
%
Much current theoretical work is devoted to finding a more fundamental
and general underlying theory from which the standard model of
particle physics could be {\hbox{deduced}} as the low energy limit.
String theory is a central candidate, with the feature that 
the assumptions underlying the CPT theorem are invalid for strings,
which are extended objects. CPT violation then becomes a possible
signature of more fundamental underlying theories, which can be probed by
sensitive experimental tests of CPT invariance.

Some years ago Kosteleck\a'{y} and coworkers 
\cite{c_hugh_kos1,c_hugh_kos2,c_hugh_kos3} developed a plausible
extension of the standard model based on spontaneous breaking of
Lorentz and CPT symmetry in an underlying theory without gravity. 
Their low energy effective theory provides a
theoretical basis for establishing quantitative bounds on CPT invariance.
The analysis was
done in the context of conventional relativistic quantum mechanics and
quantum field theory in four dimensions, retaining the usual gauge
structure and renormalizability. The Lorentz and CPT violating additions 
to the standard model Lagrangian are highly
suppressed to remain compatible with established experimental bounds.
 
As applied to muonium ($\mu^{+}e^{-}$ bound state), the most relevant
terms in the extension to the standard model are those
in the QED limit, involving only muons, electrons and photons. The 
additional terms in the Lorentz-violating Lagrangian lead to a modified 
Dirac equation and are given by \cite{c_hugh_kos2}
\begin{eqnarray}
{\mathcal{L}}&=&-a^{l}_{\alpha}\bar\psi_{l}\gamma^{\alpha}\psi_{l}
                -b^{l}_{\alpha}\bar\psi_{l}\gamma_{5}\gamma^{\alpha}\psi_{l}\\
             &~&-\frac{1}{2}H_{\alpha\beta}
                \bar\psi_{l}\sigma^{\alpha\beta}\psi_{l}+
                \frac{1}{2}ic_{\alpha\beta}
                \bar\psi_{l}\gamma^{\alpha}{\stackrel{\leftrightarrow}{D}}
                 \,^{\beta}\psi_{l}\\
             &~&+\frac{1}{2}id_{\alpha\beta}
                \bar\psi_{l}\gamma_{5}\gamma^{\alpha}
                {\stackrel{\leftrightarrow}{D}}\,^{\beta}\psi_{l}.
\end{eqnarray}
The lepton fields are denoted by $l=e^{-},\mu^{-}$, and
$iD_{\alpha}=i\partial_{\alpha}-qA_{\alpha}$ where $q=-|e|$. All terms
are Lorentz-violating, while
$a$ and $b$ are CPT odd, and $H,~c$ and $d$ are CPT even. 
The lepton number violating terms allowed in the extension are not 
relevant here and have been omitted.

\begin{figure}
\centering
\epsfxsize=\linewidth
\epsfbox{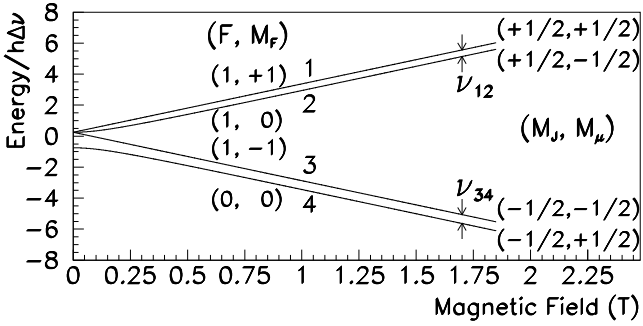}
\caption{Breit-Rabi energy level diagram of ground state muonium. At
high fields, the indicated transitions, $\nu_{12}$ and $\nu_{34}$ are 
essentially muon spin flip transitions.}
\label{fig:91breitrabi}
\end{figure}

To predict the perturbations to muonium, 
we consider the energy level diagram of the ground state
of muonium in a magnetic field (Fig. \ref{fig:91breitrabi}). The transition
frequencies $\nu_{12}$ and $\nu_{34}$ have been measured \cite{c_hugh_liu} with
high precision and are used to determine the hyperfine structure interval
$\Delta\nu$ and the ratio $\mu_{\mu}/\mu_{p}$ of the muon magnetic moment
to the proton magnetic moment.

Leading-order Lorentz violating energy shifts
$\delta\nu_{12}$ and $\delta\nu_{34}$ 
due to the new terms in the Lagrangian can be calculated
using perturbation theory and relativistic two-fermion techniques.
For our observed transitions at the strong magnetic field of 1.7 T, dominantly
only muon spin flip occurs so these energy shifts are
characterized by the muon parameters alone of the extended theory.
This approach results in :
\begin{eqnarray}
\delta\nu_{12}~\approx~-\delta\nu_{34}~
\approx~\tilde{b}_{3}^{\mu}/\pi,
\label{eqn:91four}
\end{eqnarray}
where $\tilde{b}_{3}^{\mu}~\equiv~b_{3}^{\mu}~+~d_{30}^{\mu}m_{\mu}~+~
H_{12}^{\mu}$\cite{c_hugh_bluhm} are laboratory frame parameters.
High precision experiments on muonium ($M$) can measure or set limits on the 
parameters of these symmetry violating terms, which are sensitive at
the Planck scale level \cite{c_hugh_bluhm}.

Predictions of the values of $\nu_{12}$ and $\nu_{34}$ from standard 
theory - dominantly the QED terms - requires values for many atomic
constants including $m_{\mu}$, $\mu_{\mu}$, and $\Delta\nu$ as well as the 
calculation of higher order QED radiative corrections. The relevant 
constants and calculations are not known to as high accuracy as the 
experimental determinations of $\nu_{12}$ and
$\nu_{34}$. Indeed several of these constants - $\Delta\nu$ and
$\mu_{\mu}/\mu_{p}$ - are obtained from the muonium experiment. Comparing
predictions for $\nu_{12}$ and $\nu_{34}$ (based on independent determinations
of the required atomic constants) with the experimental results results in
poor sensitivity to the non-standard model energy shifts
$\delta\nu_{12}$ and $\delta\nu_{34}$.

However, the theory with CPT and Lorentz violation involves spatial 
components in a celestial frame of reference, and since the laboratory
rotates with the earth, these spatial components vary with time, and
consequently the experimentally observed
$\nu_{12}$ and $\nu_{34}$ may oscillate about a mean value
at the earth's sidereal frequency 
$\Omega~=~2\pi/23~{\mathrm{hr}}~56~{\mathrm{m}}$ with amplitudes 
$\delta\nu_{12}$ and $\delta\nu_{34}$. No such signal would be
obtained from the standard model. In the non-rotating 
celestial frame of reference with equatorial axes 
$\big\{\hat X,\hat Y,\hat Z \big\}$ where $\hat Z$ is oriented along the 
earth's 
rotational North Pole, an experimental constraint on $\delta\nu_{12}$ 
implies\cite{c_hugh_bluhm}
\begin{eqnarray}
{\frac{1}{\pi}}|{\mathrm{sin}}\chi|
\sqrt{ \big({\tilde{b}_{X}^{\mu}}\big)^2+
       \big({\tilde{b}_{Y}^{\mu}}\big)^2} &\leq& \delta\nu_{12}
\end{eqnarray} 
in which $\chi\sim~90^{\circ}$ is the angle between $\hat Z$ and the 
quantization axis defined by the laboratory magnetic field at 
Los Alamos where the muonium experiment was performed.
The transformation from the lab frame quantity $\tilde{b}_{3}^{\mu}$ to
the non-rotating celestial frame quantities $\tilde{b}_{J}^{\mu}$ 
(where $J=X,Y,Z$) is given by
\begin{eqnarray}
\tilde{b}_{3}^{\mu}~=~\tilde{b}_{Z}^{\mu}\cos\chi~+~\big(
\tilde{b}_{X}^{\mu}\cos\Omega t ~+~\tilde{b}_{Y}^{\mu}\sin\Omega t
\big)\sin\chi.
\end{eqnarray}
The non-rotating frame quantities, $\tilde{b}_{J}^{\mu}$ are defined by
$\tilde{b}_{J}^{\mu}\equiv b_{J}^{\mu}+m_{\mu}d_{J0}^{\mu}+ 
{1\over2} \epsilon_{JKL}H_{KL}^{\mu}$ \cite{c_hugh_bluhm}.
The experiment has poor sensitivity to the celestial frame parameter
$\tilde{b}_{Z}^{\mu}$, but ideal sensitivity to the pair
$\tilde{b}_{X}^{\mu}$ and $\tilde{b}_{Y}^{\mu}$. 

The sum of the transition frequencies, $\nu_{12}+\nu_{34}$, is
equal to the ground state hyperfine splitting, $\Delta\nu$, and since we 
expect (see Eqn.\ref{eqn:91four})
$\delta\nu_{12}+\delta\nu_{34}\approx~0$, no sidereal variation is
expected in the hyperfine interval. At the high field strengths 
of this experiment, the 
difference in transition frequencies $\nu_{12}-\nu_{34}$ is
almost proportional to the magnetic moment of the muon.
By looking for a sidereal variation in the difference
$\delta\nu_{12}-\delta\nu_{34}\approx~2\tilde{b}_{3}^{\mu}/\pi$, 
we are essentially probing a possible sidereal variation of the magnetic 
moment of the muon (while the g factor remains constant).

\begin{figure}
\centering
\epsfxsize=\linewidth
\epsfbox{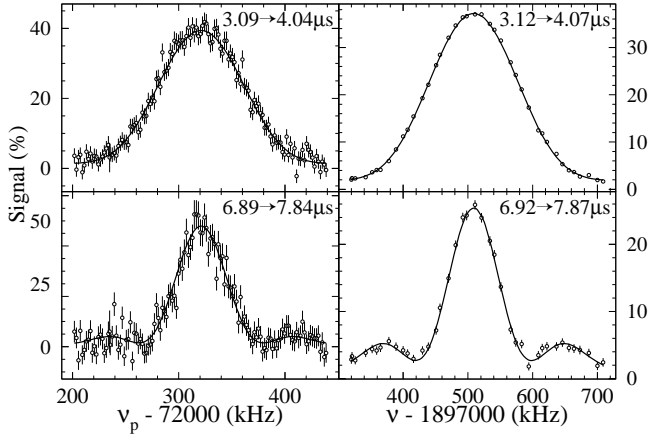}
\caption{Muonium resonance lines (data and fit) for $\nu_{12}$ taken using
magnetic field sweep on the left, and microwave frequency sweep on 
the right, for muonium atoms which have decayed in selected time
intervals after formation.}
\label{fig:91lines}
\end{figure}

The accurate measurements of $\nu_{12}$ and $\nu_{34}$ were done in
a microwave magnetic resonance experiment \cite{c_hugh_liu}. 
Muonium was formed by electron capture by muons stopping in a krypton gas 
target. Resonance lines
were observed by varying the magnetic field with fixed microwave
frequency and by varying the microwave frequency with fixed magnetic
field. A line narrowing technique was used involving observation
of a transition signal only from $M$ atoms which have lived
considerably longer than $\tau_{\mu}~\sim~2.2~\mu$s 
(Fig. \ref{fig:91lines}). The values reported for
$\nu_{12}$ and $\nu_{34}$ at a magnetic field strength corresponding to a
free proton precession frequency of $72.320~000$ MHz were 
\begin{eqnarray}
\label{eqn:91ournu12}
\nu_{12}({\mathrm{exp}})&=&1~897~539~800~(35)~{\mathrm{Hz}}~(18~{\mathrm{ppb}}),\\
\label{eqn:91ournu34}
\nu_{34}({\mathrm{exp}})&=&2~565~762~965~(43)~{\mathrm{Hz}}~(17~{\mathrm{ppb}})
\label{eqn:91ourdnu}
\end{eqnarray}
in which one standard deviation errors include both statistical and
systematic errors.
To search for a time dependence of $\nu_{12}$ and $\nu_{34}$, we have
employed the following algorithm. Data from each resonance line run 
(each lasting about half an hour) are fit at the 
measured magnetic field strength and Kr pressure to determine 
provisional line centers for $\nu_{12}$ and $\nu_{34}$. These line
centers were then transformed to their values in a magnetic field strength
corresponding to a free proton precession frequency of 72.320 000 MHz.
The data were taken at Kr pressures of 0.8 and 1.5 atm, so the line centers
were corrected for a small quadratic pressure shift, and then were 
extrapolated linearly to their values at zero pressure, using a 
pressure shift coefficient determined from the data. Line centers
for $\nu_{12}$ and $\nu_{34}$ obtained in this way were then grouped
as a function of sidereal time, where time zero has been
set as the time in 1995 when we obtained our first data.  

Non-zero values for $\delta\nu_{12}$ and $\delta\nu_{34}$ could
arise from systematic effects which lead to 
variations in the parameters affecting the line centers - particularly
variations having a period of $\approx$ 24 hr. Principal concerns are
possible day-night variations of the magnetic field strength, and
of the density and temperature of the Kr gas.

Day-night variations of several $^\circ$C
in the temperature of the experimental hall lead to oscillations in the
magnetic field strength of the persistent-mode superconducting solenoid of
$\leq$ 1 ppm. Changes of 0.05 ppm 
in the field strength were easily resolved, and the oscillation's  
effects on the line centers were accounted for in extracting the
line centers. Temperature
changes also affected the diamagnetic shielding constant of the water in the
NMR probes used to monitor the field
(the probes were not temperature-stabilized, but were in good thermal 
contact with 
the microwave cavity which was temperature-stabilized to 0.1~$^\circ$C). 
The maximum conceivable 2~$^\circ$C day-night changes in water temperature
would change the NMR frequencies by 0.02 ppm, leading to errors in the
line centers of about 2.5 Hz; of opposite sign for $\nu_{12}$ and
$\nu_{34}$. This potential effect is well below the statistical sensitivity
for sidereal variations of 12 to 15 Hz. 

The effect on these data of the variation of Kr pressure with time has been
evaluated. The front end window to the Kr stopping target was of 3 mil mylar, 
and flexed with day-night variations of the external atmospheric pressure.
This induced fractional day-night oscillations in the Kr gas target 
pressure which were measured to be 
about $2.5 \times 10^{-4}$.
Through pressure shift coefficients
of about -16.5 kHz/atm for $\nu_{12}$ and -19.5 kHz/atm for $\nu_{34}$, 
the resulting shifts in the line centers (typically 
7.5 Hz in $\nu_{34}$ and 6 Hz in $\nu_{12}$) were automatically accounted 
for in performing the linear extrapolation to zero pressure, and should not
contribute any significant time variation to $\nu_{ij}$.
The pressure shift coefficients depend on the average velocities of the 
atoms, and so are functions of
temperature. The fractional changes in the transition frequencies with
temperature (measured in hydrogen and its isotopes\cite{c_hugh_morgan}) 
are roughly 
$1\times 10^{-11}~^{\circ}{\mathrm{C}}^{-1}{\mathrm{Torr}}^{-1}$. Given the
temperature stability of the Kr gas of about 0.1~$^\circ$C, 
temperature dependent errors in the extrapolation of the line centers 
to their vacuum values would be limited to a few Hz, well below the 
statistical sensitivity of our test.

Other potential concerns involve the two frequency references used
in the experiment - the proton precession frequency forming the
basis of the magnetic field determination, and the Loran-C 10 MHz
frequency reference used for the NMR and microwave frequency synthesizers. 
The Loran-C standard is based on hyperfine transitions in Cs with $m_F$=0,
and so is insensitive to any preferred spatial orientation, and would 
not introduce a signature for Lorentz violation into the spectroscopic
measurements. Time dependent gravitational red or blue shifts of the 
frequency standard, caused by tidal distortions of the earth, are 
below the $10^{-16}$ level and too small to be seen.
Finally, bounds on clock comparisons of 
$^{199}$Hg and $^{133}$Cs \cite{c_hugh_clock,c_hugh_berglund} 
place limits on the Lorentz violating energy shifts in the precession
frequency of a proton of $10^{-27}$ GeV, which
imply the NMR measurements are free of shifts well below the Hz level.

\begin{figure}
\centering
\epsfxsize=\linewidth
\epsfbox{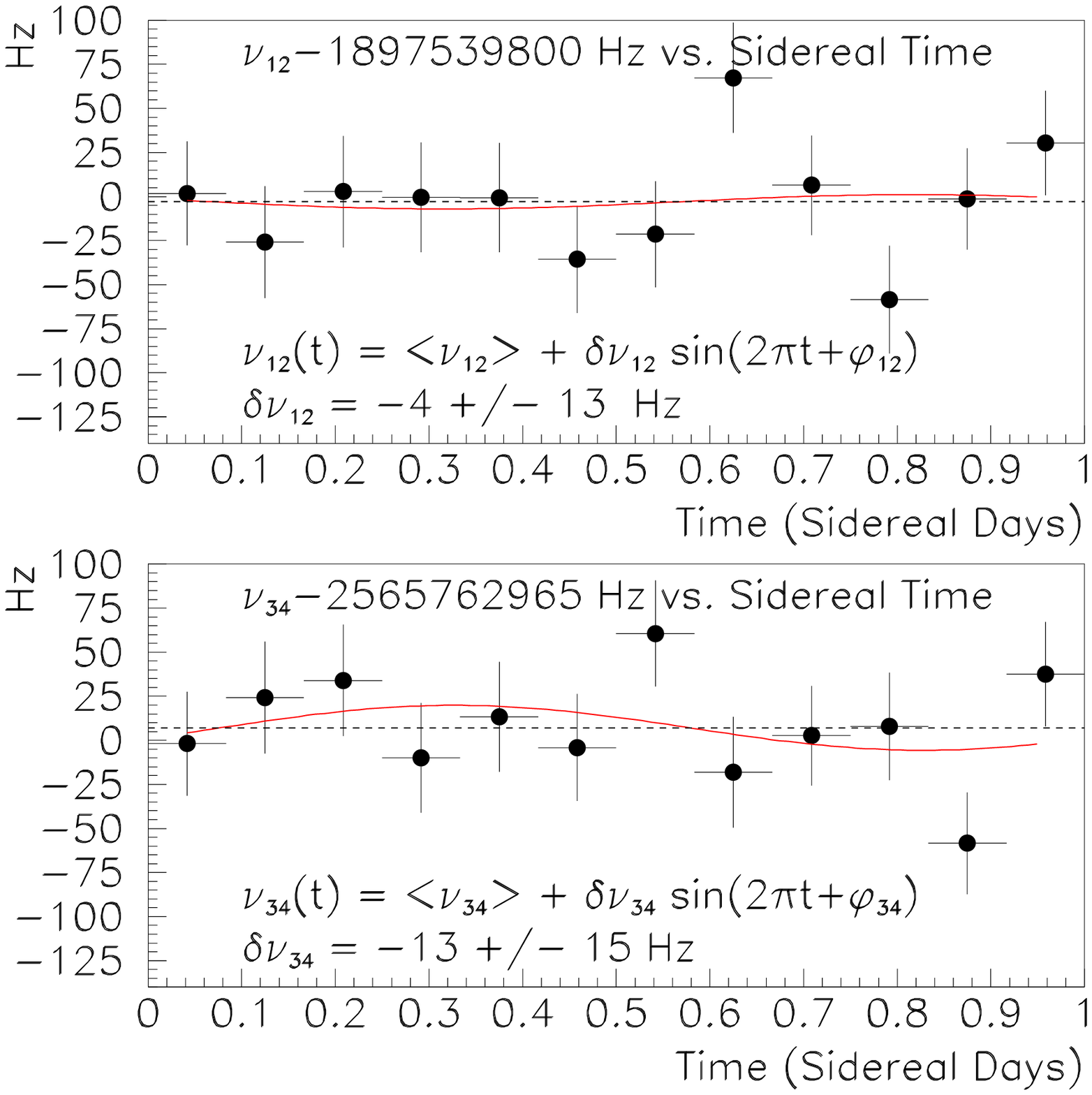}
\caption{Two years of data on $\nu_{12}$ and $\nu_{34}$ are shown binned 
versus sidereal time and fit for a possible sinusoidal variation. The
amplitudes are consistent with zero.}
\label{fig:91nuijvstime}
\end{figure}

All the data obtained in 1995 and 1996 are plotted as a function of 
time measured as a function of a sidereal day in Fig. \ref{fig:91nuijvstime},
where twelve points at $\approx$ 2 hr. intervals are plotted, and the 
vertical scale is in Hz. The data for $\nu_{12}$ and $\nu_{34}$ 
were fit separately by the functions
\begin{eqnarray}
\nu_{ij}(t)=\langle\nu_{ij}\rangle+\delta\nu_{ij}\sin
\left(2\pi t+\phi_{ij}\right)
\label{eqn:funcform}
\end{eqnarray}
where $t$ is the time in sidereal days, and the 
fit parameters are $\langle\nu_{ij}\rangle$, 
the amplitude of the possible time 
variation $\delta\nu_{ij}$, and the
phase $\phi_{ij}$ (where no phase relation was assumed between 
$\nu_{12}$ and $\nu_{34}$).
The amplitudes for $\delta\nu_{12}$ and $\delta\nu_{34}$ 
are consistent with zero, $-4\pm 13$ Hz and $-13\pm 15$ Hz
respectively.

\begin{figure}
\centering
\epsfxsize=\linewidth
\epsfbox{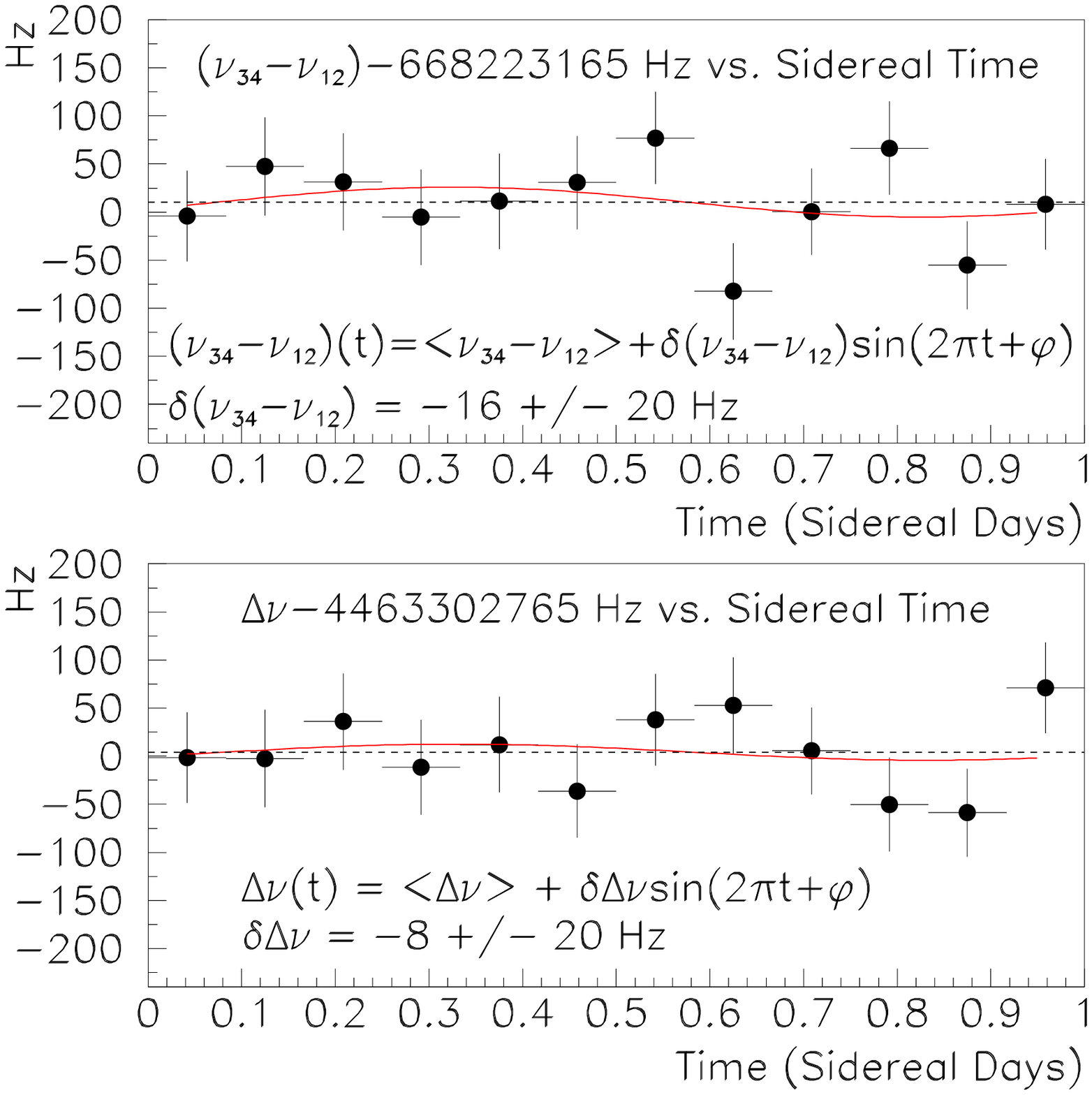}
\caption{Two years of data on $\nu_{12}$-$\nu_{34}$ and
$\nu_{12}+\nu_{34}=\Delta\nu$ are shown binned 
versus sidereal time and fit for a possible sinusoidal variation. The
amplitudes are consistent with zero.}
\label{fig:91deltavstime}
\end{figure}

As stated above, the theory being tested requires 
$\delta\nu_{12}~\approx~-\delta\nu_{34}$. A plot of $\nu_{12}-\nu_{34}$ versus
sidereal time is shown in Figure \ref{fig:91deltavstime}, and fit for
a sinusoidal variation (as in Eqn.\ref{eqn:funcform}) where a common
phase is assumed between $\nu_{12}$ and $\nu_{34}$. The data exhibit no 
variation with time within $\pm~20~$Hz, which corresponds to a
68$\%$ confidence level (one sigma) limit on the non-rotating frame components 
(see Eqn. \ref{eqn:91four})
\begin{eqnarray}
\sqrt{ \big({\tilde{b}_{X}^{\mu}}\big)^2+
       \big({\tilde{b}_{Y}^{\mu}}\big)^2} &\leq& 
       2\times 10^{-23}~{\mathrm{GeV}}.  
\end{eqnarray} 
The figure of merit of these results as a test of CPT violation is taken as 
\begin{eqnarray}
2\sqrt{ \big({\tilde{b}_{X}^{\mu}}\big)^2+
       \big({\tilde{b}_{Y}^{\mu}}\big)^2 }/{m_{\mu}}
&~{\stackrel{<}{\sim}}~& 
       \frac{2\pi\mid\delta\nu_{12}\mid}{m_{\mu}}~\approx~5\times 10^{-22}.
\end{eqnarray}
The limits on $\delta\nu_{34}$ and 
$\delta(\nu_{34}-\nu_{12})$ yield similar values.   
Other choices for the denominator in the figure of merit such as 
$\nu_{ij}$ or $\Delta\nu$ are
inappropriate since the former are B field dependent, and the latter is
not a fundamental property of the muon. It is natural to take 
the muon mass as the basic energy scale since it is an input parameter
of the standard model from which the atomic properties are derivable. The
ratio of Lorentz and CPT violating
deviations of muonium energy levels to the muon mass might be
expected to be suppressed by the ratio of the 
low-energy (standard model) scale to the Planck scale of the underlying
theory \cite{bluhmkosrus}.
In this context
the result achieved here, $5\times 10^{-22}$, is comparable to the 
dimensionless scaling factor $m_{\mu}/M_{P}~\sim~10^{-20}$.

Bounds on a different linear combination of the muon parameters 
$b_{J}^{\mu}, d_{J0}^{\mu}$, and $H_{KL}^{\mu}$ than from muonium can also
be obtained from precision measurements of the anomalous magnetic moment 
of the muon $a_{\mu}=(g_{\mu}-2)/2$ 
\cite{c_hugh_bluhm}, where one can search for sidereal variations
of the anomalous precession frequency $\omega_{a}^{\mu}$ (the difference 
between spin and cyclotron frequencies). 
Also, the parameter $\tilde{b}_{Z}^{\mu}$, which is not
tested in the latest muonium experiment, is bound at the level of 10$^{-22}$
GeV by the measurements of the anomalous precession 
frequencies $\omega_{a}^{\mu}$ of $\mu^+$ and $\mu^-$  
measured at CERN and BNL\cite{c_hugh_bluhm,c_hugh_gm2}.
Further improvements in the Lorentz and CPT violating parameters for the
muon, coming from muonium, will require higher intensity muon sources, 
as the uncertainties are predominantly statistical.

In conclusion, no unambiguous violation of CPT or Lorentz invariance has been 
observed although tests have been done on many systems since 
1960 \cite{hughesdrever}. The results presented here represent Planck 
scale sensitivity, and are an order of 
magnitude improvement in sensitivity over
previous limits on the muon parameters of the theory. 
They also represent the first test for sidereal variations in the 2nd 
generation of leptons, where Lorentz and CPT violating effects may be 
enhanced\cite{c_hugh_bluhm}.
Higher sensitivity tests as well as tests on 
different physical systems will surely be made \cite{c_hugh_kos3,coleman}.

We thank V.A. Kosteleck\a'{y} for helpful discussions, and the
U.S. DOE and BMBF (Germany) for supporting this research.
%
%

%
%
%
%
%
%
%
%
%
%
\end{document}